\documentclass[preprint,preprintnumbers,amsmath,amssymb,floatfix,nofootinbib]{revtex4}
\usepackage{graphicx}
\usepackage{url}
\usepackage{subfigure}
\newcommand{\sss}{\scriptscriptstyle}

\newcommand{\Zbb}{Z b{\bar b}}
\newcommand{\as}{\alpha_s}

\newcommand{\Zvert}[1]{\gamma^#1(g_{\sss V}^f-g_{\sss A}^f\gamma_5)}

\begin{document}          
\preprint{FSU-HEP-2008-0531}
%
\title{NLO QCD corrections to $Zb \bar b$ production with massive
bottom quarks at the Fermilab Tevatron}
\author{F.~Febres~Cordero}
\email{ffebres@physics.ucla.edu}
\affiliation{Department of Physics and Astronomy, UCLA, Los Angeles, CA 90095-1547, USA}
\author{L.~Reina}
\email{reina@hep.fsu.edu}
\affiliation{Physics Department, Florida State University,
Tallahassee, FL 32306-4350, USA}
\author{D.~Wackeroth}
\email{dow@ubpheno.physics.buffalo.edu}
\affiliation{Department of Physics, SUNY at Buffalo,
Buffalo, NY 14260-1500, USA}

\date{\today}

\begin{abstract}
We calculate the Next-to-Leading Order (NLO) QCD corrections to $Z
b\bar{b}$ production in hadronic collisions including full
bottom-quark mass effects. We present results for the total cross
section and the invariant mass distribution of the bottom-quark jet
pair at the Fermilab Tevatron $p\bar p$ collider.  We perform a
detailed comparison with a calculation that considers massless bottom
quarks, as implemented in the Monte Carlo program MCFM.  We find that
neglecting bottom-quark mass effects overestimates the total NLO QCD
cross section for $Zb\bar{b}$ production at the Tevatron by about 7\%,
independent of the choice of the renormalization and factorization
scales. Moreover, bottom-quark mass effects can impact the shape of
the bottom-quark pair invariant mass distribution, in particular in
the low invariant mass region.
\end{abstract}
%
\maketitle

\section{Introduction}
\label{sec:intro}
One of the main goals of high-energy collider experiments is the
elucidation of the mechanism of Electroweak Symmetry Breaking (EWSB)
as well as the exploration of energy scales beyond the weak scale,
where physics beyond the Standard Model (SM) is expected.  The
hadronic production of weak gauge bosons in association with a
bottom-quark pair plays a crucial role in some of the current studies
of EWSB and beyond the SM physics at the Fermilab Tevatron $p\bar p$
collider~\cite{Acosta:2005ga,Abulencia:2005ep,Aaltonen:2007wx,Abazov:2004jy,Abazov:2007hk,Abazov:2007pj}. $W
b\bar b$ and $Zb\bar b$ production processes represent the major
irreducible backgrounds to the main search modes for a light SM-like
Higgs boson at the Tevatron, i.e.~$WH$ and $ZH$ with $H\to b\bar
b$. $Wb\bar{b}$ also accounts for one of the most important
backgrounds to single-top production, $p\bar{p}\rightarrow
t\bar{b},\bar{t}b$ with $t(\bar{t})\rightarrow Wb(\bar{b})$, which
tests the fundamental structure of the $Wtb$ vertex at the Tevatron
\cite{Acosta:2004bs,Abazov:2005zz,CDFnote:2008st,Abazov:2008kt}. Finally,
$Zb\bar b$ is a background to searches for Higgs bosons in models with
enhanced bottom-quark Yukawa couplings, such as the Minimal
Supersymmetric Standard Model (MSSM) with large $\tan\beta$, where
$Hb\bar b$ with $H \to \mu^+ \mu^-,\tau^+\tau^-$ is an interesting
discovery channel~\cite{Kao:2007qw}.

The hadronic cross sections for $W/ZH$ associated production have been
calculated including up to Next-to-Next-to-Leading Order (NNLO) QCD
corrections~\cite{Han:1991ia,Mrenna:1997wp,Brein:2003wg} and
$O(\alpha)$ electroweak corrections~\cite{Ciccolini:2003jy}.
Single-top production has been calculated at Next-to-Leading (NLO) in
QCD~\cite{Stelzer:1997ns,Stelzer:1998ni,Smith:1996ij,Harris:2002md,Sullivan:2004ie,Cao:2004ky,Cao:2004ap,Cao:2005pq,Sullivan:2005ar},
and including one-loop electroweak (SM and MSSM)
corrections~\cite{Beccaria:2006ir}, while the cross section for
$Hb\bar{b}$ associated production is known including NLO QCD
corrections and full bottom-quark mass
effects~\cite{Dittmaier:2003ej,Dawson:2003kb,Dawson:2004sh,Dawson:2004wq}.

To fully exploit the Tevatron's potential to detect the SM Higgs boson
or to impose limits on its mass, it is crucial that the dominant
background processes are also under good theoretical control.  In the
present experimental analyses~\footnote{For updated results, see the
CDF and $D\emptyset$ websites at
http://www-cdf.fnal.gov/physics/exotic/exotic.html and
http://www-d0.fnal.gov/Run2Physics/WWW/results/higgs.htm.}, the
effects of NLO QCD corrections on the total cross section and the
dijet invariant mass distribution of the $W/Z\ b\bar{b}$ background
processes have been taken into account by using the Monte Carlo
program MCFM~\cite{MCFM:2004}. In MCFM, the NLO QCD predictions of
both total and differential cross sections for these processes have
been calculated in the zero bottom-quark mass ($m_b=0$)
approximation~\cite{Ellis:1998fv,Campbell:2000bg,Campbell:2002tg},
using the analytical results of Refs.~\cite{Bern:1996ka,Bern:1997sc}.
From a study of the Leading Order (LO) cross section, finite
bottom-quark mass effects are expected to affect both the total and
differential $W/Zb\bar{b}$ cross sections, mostly in the region of
small $b\bar{b}$-pair invariant masses~\cite{Campbell:2002tg}. Indeed,
since this kinematic region of small $b\bar{b}$-pair invariant masses
contributes considerably to $W/Z+n\,j$ production ($n=1,2$), where at
least one of the jets is a $b$-jet, bottom-quark mass effects cannot
be neglected as discussed in
Refs.~\cite{Campbell:2006cu,Campbell:2005zv} (for $n=2$) and in
Ref.~\cite{Campbell:2008hh} (for $n=1$).  Given the variety of
experimental analyses involved in the search for $W/ZH$ associated
production, single-top and $Hb\bar{b}$ production, it is important to
assess precisely the impact of a finite bottom-quark mass over the
entire kinematical reach of the process, including the complete NLO
QCD corrections.

In Ref.~\cite{FebresCordero:2006sj} we have performed a study of NLO
QCD cross sections and invariant mass distributions of the bottom
quark pair in $W b \bar b$ production at the Tevatron, including full
bottom-quark mass effects. We found that bottom-quark mass effects
amount to about 8\% of the total NLO cross section at the Tevatron and
are mostly visible in the region of low $b\bar{b}$-pair invariant
mass.

In this paper, we compute the NLO QCD corrections to $Zb\bar b$
hadronic production, including the full bottom-quark mass effects. We
consider all partonic processes that contribute at ${\cal O}(\alpha
\alpha_s^3)$, i.e.~NLO QCD corrections to $q\bar q \to Z b\bar b$ and
$gg \to Zb\bar b$ and the tree-level process $q(\bar q)g \to Zb\bar b
q(\bar q)$.  We present numerical results for the total cross section
and the invariant mass distribution of the $b\bar{b}$ pair for the
Tevatron $p\bar{p}$ collider, including kinematic cuts and a
jet-finding algorithm. In particular, we apply the $k_T$ jet algorithm
and require two tagged $b$-jets in the final state.  Using the MCFM
package~\cite{MCFM:2004}, we compare our results with the
corresponding results obtained in the $m_b=0$ limit. Numerical results
for both $Zb\bar{b}$ and $Wb\bar{b}$ production at the Large Hadron
Collider will be presented in a separate publication~\cite{FebresCordero:2008lhc}.

The outline of the paper is as follows: in
Section~\ref{sec:calculation} we briefly discuss the technical details
of our calculation, while we present numerical results and a
discussion of the bottom-quark mass effects in
Section~\ref{sec:results}. Section~\ref{sec:conclusions} contains our
conclusions.
\begin{figure}[htb]
\begin{center}
\includegraphics[scale=0.75]{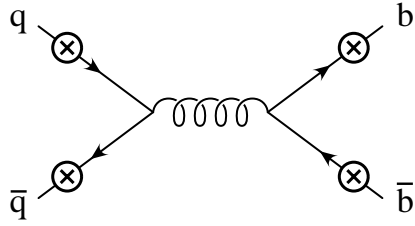}
\caption{Tree-level Feynman diagrams for $q\bar{q}\to\Zbb$. The circled crosses indicate all possible
insertions of the final-state $Z$ boson leg, each insertion corresponding to a different diagram.}
\label{fig:qqZbb_tree}
\end{center}
\end{figure}
\begin{figure}[htb]
\begin{center}
\includegraphics[scale=0.9]{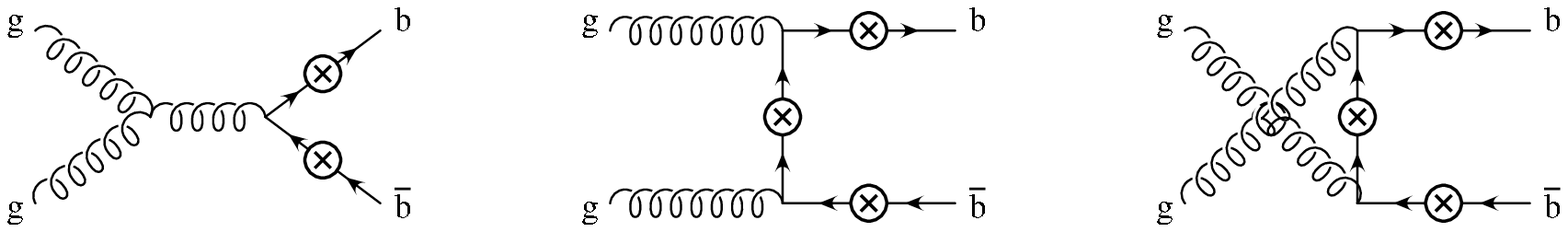}
\caption{Tree-level Feynman diagrams for $gg\to\Zbb$. The circled crosses indicate all possible
insertions of the final-state $Z$ boson leg, each insertion corresponding to a different diagram.}
\label{fig:ggZbb_tree}
\end{center}
\end{figure}
\section{Calculation}
\label{sec:calculation}
\subsection{Basics}
\label{subsec:basics}
The total cross section for $p\bar{p}(pp)\to Zb\bar{b}$ at ${\cal
  O}(\alpha \alpha_s^3)$ can be written as follows:
\begin{eqnarray}
\sigma^{\rm NLO}(p{\bar p}(pp)\to Z b{\bar b}) 
&=& \sum_{ij}\frac{1}{1+\delta_{ij}} \int dx_1dx_2 \nonumber\\
& &  \qquad \left[ {\cal F}_i^p(x_1,\mu) {\cal F}_j^{{\bar p(p)}}(x_2,\mu) 
{\hat\sigma}_{ij}^{\rm NLO}(x_1,x_2,\mu) + (x_1\leftrightarrow x_2) \right],
\label{eq:HaddXsec}
\end{eqnarray}
where ${\cal F}_i^{p({\bar p})}$ denote the parton distribution
functions (PDFs) for parton $i$ in a proton (antiproton), defined at a
generic factorization scale $\mu_f=\mu$.  The sum runs over all
relevant subprocesses contributing to the hadronic cross section
initiated by partons $i$ and $j$. The partonic cross section for the
subprocess $ij\to Z b{\bar b}\,\,(+k)$ is denoted by
${\hat\sigma}_{ij}^{\rm NLO}$ and is renormalized at an arbitrary
scale $\mu_r=\mu$. If not specified otherwise, we assume the
factorization and renormalization scales to be equal,
$\mu_f=\mu_r=\mu$.  The factor in front of the integral is a symmetry
factor that accounts for the presence of identical particles in the
initial state of a given subprocess ($\delta_{ij}$ is the Kronecker
delta).  The partonic center-of-mass energy squared, $s$, is given in
terms of the hadronic center of mass energy squared, $s_H$, by
$s=x_1x_2s_H$.

The NLO QCD partonic cross section reads:
\begin{eqnarray}
{\hat\sigma}_{ij}^{\rm NLO}(x_1,x_2,\mu) & = & {\hat\sigma}_{ij}^{\rm LO}(x_1,x_2,\mu)
				+\delta {\hat\sigma}_{ij}^{\rm NLO}(x_1,x_2,\mu)\; ,
\label{eq:nloXsec}
\end{eqnarray}
where ${\hat \sigma}_{ij}^{\rm LO}(x_1,x_2,\mu)$ denotes the ${\cal
  O}(\alpha \alpha_s^2)$ LO partonic cross section and $\delta
{\hat\sigma}_{ij}^{\rm NLO}(x_1,x_2,\mu)$ describes the ${\cal
  O}(\alpha_s)$ corrections to ${\hat \sigma}_{ij}^{\rm
  LO}(x_1,x_2,\mu)$.  The LO $p\bar p(pp)\to Z b \bar b$ process
receives contributions from $q\bar q$ and $gg$ initiated processes, as
shown in Fig.~\ref{fig:qqZbb_tree} and
Fig.~\ref{fig:ggZbb_tree}, respectively.  The NLO QCD corrections, $\delta
{\hat\sigma}_{ij}^{\rm NLO}$, receive contributions from $q\bar q$,
$gg$, $qg$ and $\bar q g$ initiated processes and can be decomposed in
the following way:
\begin{eqnarray}
\delta {\hat\sigma}_{ij}^{\rm NLO} & = & \int d\left( PS_3\right) {\overline \sum}|
	{\cal A}_{\rm virt}(ij\to Zb\bar b)|^2 + \int d\left( PS_4\right) {\overline \sum}|
	{\cal A}_{\rm real}(ij\to Zb \bar b +k)|^2 \nonumber\\
&\equiv & {\hat\sigma}_{ij}^{\rm virt} +{\hat\sigma}_{ij}^{\rm real}\;,
\label{eq:virtrealXsec}
\end{eqnarray}
where the term integrated over the phase space measure
$d\left(PS_3\right)$ corresponds to the virtual one-loop corrections
with three particles in the final state, while the one integrated over
the phase space measure $d\left(PS_4\right)$ corresponds to the real
tree-level corrections with one additional emitted parton.  The sum
$\overline \sum$ indicates that the corresponding amplitudes squared,
$|{\cal A}_{\rm virt(real)}(ij\to Zb \bar b(+k))|^2$, have been
averaged over the initial-state degrees of freedom and summed over the
final-state ones. The phase space integration has been performed using
Monte Carlo (MC) techniques using the adaptive multi-dimensional
integration routine VEGAS~\cite{Lepage:1977sw}.

We have improved on the massless calculation of
Refs.~\cite{Campbell:2000bg,Campbell:2002tg} by considering a fully
massive $b$-quark both at the level of the scattering amplitude and in
the integration over the phase space of the final-state particles.  We
keep the $Z$ boson on-shell, though the extension to include its
leptonic decays does not present in principle any special
complications.  Because of the complexity of this calculation, all
results have been cross-checked with at least two independent sets of
codes.  The analytic calculation of the scattering amplitudes has been
implemented using, at different stages, FORM~\cite{Vermaseren:2000nd},
TRACER~\cite{Jamin:1991dp}, {\it Mathematica} and {\it Maple}. Final
numerical results have been obtained with codes built in $C$ and
FORTRAN, and we have used the FF~\cite{vanOldenborgh:1990yc} and
Madgraph~\cite{Murayama:1992gi,Stelzer:1994ta,Maltoni:2002qb} packages
for cross-checks.

The ${\cal O}(\alpha_s)$ corrections to $Zb\bar{b}$ production are
similar in structure to the NLO QCD corrections to $Wb\bar{b}$
production~\cite{FebresCordero:2006sj} and to Higgs production in
association with top quarks
($Ht\bar{t}$)~\cite{Reina:2001bc,Dawson:2003zu}.  We therefore only
summarize below the most important features of the calculation and
refer to the $Wb\bar{b}$ and $Ht\bar{t}$ papers cited above and to
Ref.~\cite{Cordero:2008ce} for more details.  The ${\cal
O}(\alpha_s)$ corrections to $q\bar{q}\rightarrow Zb\bar{b}$ can be
derived from the ${\cal O}(\alpha_s)$ corrections to $q \bar{q}' \to Wb\bar{b}$
production~\cite{FebresCordero:2006sj} (with $W\leftrightarrow Z$) and
to $q\bar{q}\rightarrow Ht\bar{t}$~\cite{Reina:2001bc} (with
$t\leftrightarrow b$ and $H\leftrightarrow Z$), while the ${\cal
O}(\alpha_s)$ corrections to the $gg$ initiated $Zb\bar{b}$ production
process can be obtained from the ${\cal O}(\alpha_s)$ corrections to
$gg \to Ht\bar{t}$~\cite{Dawson:2003zu} (with $t\leftrightarrow b$ and
$H\leftrightarrow Z$). The $qg,\bar q g$ initiated processes appear in
both $Wb\bar{b}$ and $Ht\bar{t}$ NLO QCD calculations with
$W\leftrightarrow Z$ and $H\leftrightarrow Z, t\leftrightarrow b$,
respectively. Note that, when applying the results of $Wb\bar{b}$ and
$Ht\bar{t}$ production of Ref.~\cite{FebresCordero:2006sj} and
Refs.~\cite{Reina:2001bc,Dawson:2003zu}, one also needs to replace
respectively the $Wff'$ and fermion Yukawa couplings by the $V-A$
coupling of fermions to the $Z$ boson:

\begin{tabular}{rl}
\hspace{1.0cm}
\begin{minipage}{0.2\linewidth}{
\vspace{0.7cm}
\begin{center}
\includegraphics*{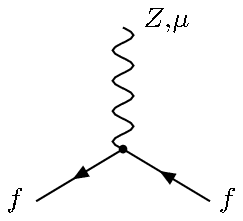}
\end{center}}
\end{minipage}
&
\hspace{-2.0cm}
\begin{minipage}{0.5\linewidth}{
\begin{center}
{\[\qquad = \;\; \frac{-i e}{2\sin{\theta_{\sss W}} \cos{\theta_{\sss W}}}\Zvert{\mu}\; ,  \]}
\end{center}}
\end{minipage}
\end{tabular}
\vspace{-1.8cm}
\begin{equation}\label{eq:FRqqZ}\end{equation}

\vspace{1.5cm}
\noindent
where $\theta_{\sss W}$ is the weak mixing angle and the vector, $g_V^f$, and axial-vector, $g_A^f$,
couplings for the $Zff$ vertex are given by:
\begin{eqnarray}
g_V^f  =  T_3^f-2\sin^2{\theta_{\sss W}} Q_f\; \; &,&
\; \;  g_A^f  =  T_3^f\; ,
\end{eqnarray}
with $T_3^f$ denoting the third component of the weak isospin 
and $Q_f$ the electric charge of the fermion $f$.
Moreover, as we are considering an on-shell $Z$ boson, we have summed over 
its polarizations as follows:
\begin{equation}
\sum \epsilon^\mu(p_Z)\epsilon^{\nu *}(p_Z)=-g^{\mu\nu}+\frac{p_Z^{\mu} p_Z^{
\nu}}{M_{\sss Z}^2}\; ,
\label{eq:polsum}
\end{equation}
where $M_{\sss Z}$ is the mass of the $Z$ boson. 
\boldmath
\subsection{The virtual cross section $\hat\sigma^{\rm virt}_{ij}$}\label{subsec:sigvirt}
\unboldmath The ${\cal O}(\as)$ virtual corrections to the partonic
tree-level $q\bar q \to \Zbb$ and $gg \to \Zbb$ production processes
consist of self-energy, vertex, box and pentagon diagrams, as shown,
for the $Ht\bar{t}$-like part, in Figs.~2-4 of
Ref.~\cite{Reina:2001bc} and Figs.~2-5 of Ref.~\cite{Dawson:2003zu},
respectively (for a full set of diagrams see also Ref.~\cite{Cordero:2008ce}).  
The contributions to $\hat\sigma^{\rm virt}_{ij}$ in
Eq.~(\ref{eq:virtrealXsec}) can be written as:
\begin{equation}
{\overline \sum}|{\cal A}_{\rm virt}(ij\to \Zbb)|^2 
	= \sum_D{\overline \sum}\left( {\cal A}_0{\cal A}_D^\dag + {\cal A}_0^\dag{\cal A}_D\right)
	= \sum_D{\overline \sum}2{\cal R}e\left( {\cal A}_0{\cal A}_D^\dag \right)\;,
\label{eq:virtXsec}
\end{equation}
where ${\cal A}_0$ is the tree level amplitude and ${\cal A}_D$
denotes the amplitude for the one-loop diagram $D$, with $D$ running
over all self-energy, vertex, box and pentagon diagrams corresponding
to the $ij$-initiated subprocess.

The calculation of each virtual diagram (${\cal A}_D$) is performed in
the same way as outlined in Refs.~\cite{Reina:2001bc,Dawson:2003zu}, i.e.
${\cal A}_D$ is calculated as a linear combination of Dirac structures
with coefficients that depend on both scalar and tensor one-loop
Feynman integrals with up to five denominators. We solve the one-loop
integrals in the coefficients either at the level of the amplitude or
at the level of the amplitude squared (see
Eq.~(\ref{eq:virtXsec})). These two independent approaches allow us to
thoroughly cross-check the calculation of each individual
diagram. Indeed, the tensor structures present in the one-loop
integrals of the amplitude are typically different from the ones
present in the amplitude squared, as one can perform non-trivial
reductions of the latter by canceling dot-products of the integration
momentum in the numerator with denominators in the Feynman
integrals. In this way, the final analytical expression of a given
diagram ends up being represented in terms of different building
blocks. A possible incorrect relation between the building blocks
would then naturally produce a discrepancy between the two approaches.

Tensor and scalar one-loop integrals are treated as follows. Using the
Passarino-Veltman (PV) method~\cite{Passarino:1978jh,Denner:1993kt},
the tensor integrals are expressed as a linear combination of tensor
structures and coefficients, where the tensor structures depend on the
external momenta and the metric tensor, while the coefficients depend
on scalar integrals, kinematics invariants and the dimension of the
integral. Numerical stability issues may arise at this level as a
consequence of the proportionality of the tensor integral coefficients
to powers of inverse Gram Determinants (GDs), as discussed in detail
in Ref.~\cite{Dawson:2003zu}, although for $Zb\bar b$ production the problem is considerably more serious.

These numerical instabilities can be considered as ``spurious'' or
``unphysical'' divergences, since it is well known that only
two-particle invariants can give rise to a physical
singularity. Indeed, these spurious divergences cancel when large sets
of diagrams are combined~\cite{Bern:1997sc}, such as, for example,
when one combines gauge invariant sets of color amplitudes
(i.e. amplitudes with a common color factor).  As we have expressed
our calculation in terms of invariants, and we employ a standard basis
of scalar integrals, the full cancellation only occurs between
numerator and denominator at the numerical level, often between fairly
large expressions.

For this reason we have chosen to organize the diagrams, at certain
stages, into gauge invariant color amplitudes, that is, into
coefficients of the same color structure (see
Ref.~\cite{Cordero:2008ce}). This allows a better handling of
the spurious singularities and a natural way to make internal
cross-checks and cross-checks with new techniques.  When we consider
these gauge invariant sets of color amplitudes and full analytical
reductions of all tensor integrals, we find cancellation of some
powers of GDs, which improves the numerical stability of our code, so
that when integrating over the $Zb \bar b$ phase space, using MC
techniques, we obtain statistical errors below 0.1\% for total cross
sections.

The fully reduced numerical codes are often more demanding
computationally, and because of that we have built master codes that
use them only when close to regions of phase space where certain
problematic GDs become small. All this is found particularly useful
when considering higher rank $D$-PV functions (we have up to $D4$-PV
functions in our calculation) as well as $E$-PV functions.  This
technique would probably break down if one were to extend it to
processes with even more legs, and the use of helicity amplitudes
would in this case be preferable.

In the case of pentagon diagrams, a powerful and convenient check
consists of reducing consistently all $E$-PV functions by canceling
systematically, at the level of the amplitude squared in
Eq.~(\ref{eq:virtXsec}), all possible vector products containing the
loop momentum in the numerator with some denominators. This is
possible as, in the pentagon topology of our process, each leg has an
outgoing momentum which is on-shell, corresponding basically to one of
the external initial or final particles of the subprocess. One then
ends with expressions for each pentagon diagram containing purely
scalar pentagon integrals, or tensor integrals with fewer than five
denominators, improving considerably the numerical stability.
We compared analytically these reductions to the non-reduced
expressions by using the full reduction of all tensor integrals to
scalar integrals, and found agreement. 

\begin{figure}[ht]
\begin{center}
\includegraphics[scale=1.0]{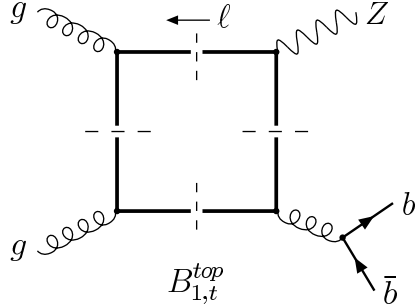}
\caption{Quadruple cut~\cite{Britto:2004nc} check of the calculation 
of a box diagram involving a top-quark loop. It corresponds to
  two Feynman diagrams given by the two possible orientations of the
  fermion line.}
\label{fig:b1ttop}
\end{center}
\end{figure}

We also checked parts of our result by using unitarity
techniques~\cite{Bern:1997sc}, specifically the quadruple-cut
technique~\cite{Britto:2004nc}. As shown by Britto, Cachazo and Feng
(BCF), from any set of Feynman diagrams (or more generally from any
tensor integral~\cite{Ossola:2006us}) one can extract the coefficient
of a given scalar box integral by cutting the four corresponding
propagators (see Fig.~\ref{fig:b1ttop}), i.e. by replacing
$i/(p^2-m^2+i\epsilon)\rightarrow 2\pi\delta^{(+)}(p^2-m^2)$ for each
cutted propagator of momentum $p$ and mass $m$. This effectively
freezes the momentum integration, and replaces it by a set of
algebraic equations which determine the loop momentum entirely.  We
solved this set of equations by using a BCF
ansatz~\cite{Britto:2004nc}, and then compared the result to the
corresponding box coefficient extracted from our analytic expression,
and found agreement (for more details and specific solutions for the
topology in Fig.~\ref{fig:b1ttop} see
Ref.~\cite{Cordero:2008ce}). This is a rather non-trivial check
for the set of $E$-PV and $D$-PV functions we have employed at different
stages, since they all contribute to the coefficients of the scalar
$D$-functions occurring in the one-loop $Zb\bar{b}$ amplitude. For
instance, it has been particularly useful in the case of box diagrams
like the one shown in Fig.~\ref{fig:b1ttop}, since this diagram and
related ones contain up to $D4$-PV functions that cannot be reduced even
at the level of the amplitude squared. Since they involve up to four
powers of inverse GDs, they are particularly subject to numerical
instabilities and it is important to have their analytic expressions
as compact as possible.

After the tensor integral reduction is performed, the fundamental
building blocks are one-loop scalar integrals with up to five
denominators. They may be finite or contain both ultraviolet (UV) and
infrared (IR) divergences. The finite scalar integrals are evaluated
using the method described in Ref.~\cite{Denner:1993kt} and
cross-checked with the numerical package
FF~\cite{vanOldenborgh:1990yc}. The UV and IR singular scalar
integrals are calculated analytically by using dimensional
regularization in $d\!=\!4-2\epsilon$ dimensions. The most difficult
integrals arise from IR divergent pentagon diagrams with several
external and internal massive particles. We calculate them as linear
combinations of box integrals using the method of
Refs.~\cite{Bern:1992em,Bern:1993kr} and of Ref.~\cite{Denner:1993kt}.
Details of the box scalar integrals (see also
Ref.~\cite{Ellis:2007qk}) and the pentagon reduction, as well as the
set of IR-divergent three and two-point functions used in this
calculation, are given in Ref.~\cite{Cordero:2008ce}.

The UV singularities of the virtual cross section are removed by
introducing a suitable set of counterterms (see
Refs.~\cite{Reina:2001bc,Dawson:2003zu,Cordero:2008ce} for details), while the
residual renormalization scale dependence is checked from first
principles using renormalization group arguments as in
Eq.~(4) of Ref.~\cite{Dawson:2003zu}.
Note that we use the on-shell subtraction scheme when fixing the wave
function renormalization constant of the external bottom quark field
($\delta Z_2^{(b)}$) and the mass renormalization constant ($\delta
m_b$).  The IR singularities of the virtual cross section are canceled by
analogous singularities in the ${\cal O}(\alpha \as^3)$ real cross
section.

In our calculation we treat $\gamma_5$ according to the naive
dimensional regularization approach, i.e.  we enforce the fact that
$\gamma_5$ anticommutes with all other $\gamma$ matrices in
$d=4-2\epsilon$ dimensions. This is known to give rise to
inconsistencies when, at the same time, the $d$-dimensional trace of
four $\gamma$ matrices and one $\gamma_5$ is forced to be non-zero (as
in $d=4$, where
$Tr(\gamma^\mu\gamma^\nu\gamma^\rho\gamma^\sigma\gamma_5)=4i\epsilon^{\mu\nu\rho\sigma}$)
~\cite{Larin:1993tq}.  In our calculation, both UV and IR divergences
are handled in such a way that we never have to enforce simultaneously
these two properties of the Dirac algebra in $d$ dimensions. For
instance, the UV divergences are extracted and canceled at the
amplitude level, after which the $d\rightarrow 4$ limit is taken and
the renormalized amplitude is squared using $d=4$. Thus, all fermion
traces appearing at this point are computed in four dimensions and
therefore have no ambiguities.

We note that the tree-level amplitude ${\cal A}_0$ in
Eq.~(\ref{eq:virtXsec}) has generically to be considered as a
$d$-dimensional tree-level amplitude. This matters when the ${\cal
  A}_D$ amplitudes in Eq.~(\ref{eq:virtXsec}) are UV or IR
divergent. Actually, as it has been shown in
Refs.~\cite{Reina:2001bc,Dawson:2003zu}, both UV and IR divergences
are always proportional to the tree level amplitudes and they can be
formally canceled without having to explicitly specify the
dimensionality of the tree level amplitude itself. After UV and IR
singularities have been canceled, 
the remaining phase space integration is computed in $d=4$
dimensions using standard MC techniques.

\boldmath
\subsection{The real cross section $\hat\sigma^{\rm real}_{ij}$}\label{subsec:sigreal}
\unboldmath The NLO QCD real cross section $\hat\sigma^{\rm
  real}_{ij}$ in Eq.~(\ref{eq:virtrealXsec}) corresponds to the ${\cal
  O}(\alpha_s)$ corrections to $ij\to \Zbb$ due to the emission of an
additional real parton, i.e. to the process $ij\to \Zbb+g$, and the
tree-level process $q(\bar q) g \to Zb\bar b+q(\bar q)$.
$\hat\sigma^{\rm real}_{ij}$ contains IR singularities which cancel
the analogous singularities present in the ${\cal O}(\alpha_s)$
virtual corrections and in the NLO PDFs (see
Refs.~\cite{Reina:2001bc,Dawson:2003zu,Cordero:2008ce} for details).  These
singularities can be either \emph{soft}, when the emitted extra parton
is a gluon and its energy becomes very small, or \emph{collinear},
when the final-state parton is emitted collinear to one of the partons
in the initial state. There is no collinear singularity arising from
the radiation off the final-state bottom quarks, since they are
considered to be massive.

We have calculated the cross sections for the processes
\[
i(q_1)+j(q_2)\to b(p_b)+\bar{b}(p_{\bar b})+Z(p_{\sss Z})+g(k)
\]
and
\[
(q,\bar q)(q_1)+g(q_2)\to b(p_b)+\bar{b}(p_{\bar b})+Z(p_{\sss Z})+(q, \bar q)(k)\,\,\, ,
\]
with $q_1+q_2=p_b+p_{\bar b}+p_{\sss Z}+k$, using the
\emph{two-cutoff} Phase Space Slicing (PSS) method.  This
implementation of the PSS method was originally developed to study QCD
corrections to dihadron production~\cite{Bergmann:1989zy} and has
since then been applied to a variety of processes (a nice review can
be found in Ref.~\cite{Harris:2001sx}). We follow closely the
application of the PSS method to $Ht\bar{t}$ production as presented in
Refs.~\cite{Reina:2001bc,Dawson:2003zu} to which we refer for more
extensive references and full details. Although we are considering $Zb\bar
b$ production, the kinematics are equivalent, and the color structure
and IR behavior are the same, so necessarily their soft and collinear
kernels are the same. In the following we briefly summarize our
implementation of the \emph{two-cutoff} PSS method.

Using the PSS method, the IR singularities can be conveniently
isolated by \emph{slicing} the phase space of the final-state
particles into different regions defined by suitable cutoffs. To
isolate the soft and collinear singularities we impose soft
($\delta_s$) and collinear ($\delta_c$) cutoffs on the phase space of
the emitted parton as follows.  By introducing an arbitrary small
\emph{soft} cutoff $\delta_s$, we separate the overall integration of
the $q\bar q, gg\to b\bar{b}Z+g$ phase space into two regions
according to whether the energy of the final state gluon
($k^0\!=\!E_g$) is \emph{soft}, i.e.  $E_g\le\delta_s\sqrt{s}/2$, or
\emph{hard}, i.e.  $E_g>\delta_s\sqrt{s}/2$. In order to isolate the
collinear singularities we further divide the hard region of the $q
\bar q, gg\to b\bar{b}Z+g$ phase space into a \emph{hard/collinear}
and a \emph{hard/non-collinear} region, by introducing a second small
\emph{collinear} cutoff $\delta_c$.  The \emph{hard/non-collinear} region
is defined by the condition that both
\begin{equation}
\label{eq:deltac_cuts}
\frac{2 q_1\!\cdot\! k}{E_g \sqrt{s}}> \delta_c\,\,\,\,\,\,\,
\mbox{and}\,\,\,\,\,\,\,
\frac{2 q_2\!\cdot\! k}{E_g\sqrt{s}}>\delta_c
\end{equation}
are true. We apply the same collinear cutoff to the tree-level process
$q(\bar q) g \to Zb\bar b+q(\bar q)$. The hard non-collinear parts of
the real cross sections, $\hat{\sigma}_{q\bar
q,gg,qg}^{hard/non-coll}$, are finite and can be computed numerically.
The partonic real cross sections can then be written as follows:
\begin{equation}
\label{eq:sigma_real_two_cutoff}
\hat{\sigma}_{q\bar q,gg,qg}^{\rm real} =
\hat{\sigma}_{q\bar q, gg}^{soft}+\hat{\sigma}_{q\bar q,gg,qg}^{hard/coll}+\hat{\sigma}_{q\bar q,gg,qg}^{hard/non-coll}\,\,\,,
\end{equation}
where $\hat\sigma_{q\bar q,gg}^{soft}$ and $\hat{\sigma}_{q\bar
  q,gg,qg}^{hard/coll}$ is obtained by integrating analytically over
the \emph{soft} and \emph{collinear} regions of the phase space of the
emitted parton, respectively, and contains all the IR divergences of
$\hat\sigma^{\rm real}_{q\bar q,gg,qg}$.  The dependence on these
arbitrary cutoffs, $\delta_s,\delta_c$, is not physical, and cancels
at the level of the real cross section, i.e. in $\hat \sigma^{\rm
  real}_{ij}$.  This cancellation constitutes an important check of
the calculation.

We conclude this section by showing explicitly that the total hadronic
cross section at NLO QCD does not depend on the arbitrary cutoffs
introduced by the PSS method, i.e. on $\delta_s$ and $\delta_c$. The
cancellation of the PSS cutoff dependence is realized in $\hat
\sigma^{\rm real}_{ij}$ by matching contributions that are calculated
either analytically ($\hat\sigma_{ij}^{soft}$ and
$\hat{\sigma}_{ij}^{hard/coll}$), in the IR-unsafe region below the
cutoffs, or numerically, in the IR-safe region above the cutoffs
($\hat \sigma_{ij}^{hard/non-coll}$).  While the analytical calculation in the
IR-unsafe region reproduces the form of the cross section in the soft
or collinear limits and is therefore only accurate for small values of
the cutoffs, the numerical integration in the IR-safe region becomes
unstable for very small values of the cutoffs. Therefore, obtaining a
convincing cutoff independence involves a delicate balance between the
previous antagonistic requirements and ultimately dictates the choice
of values that are neither too large nor too small for the cutoffs.
\begin{figure}[ht]
\begin{center}
\includegraphics*[scale=0.55]{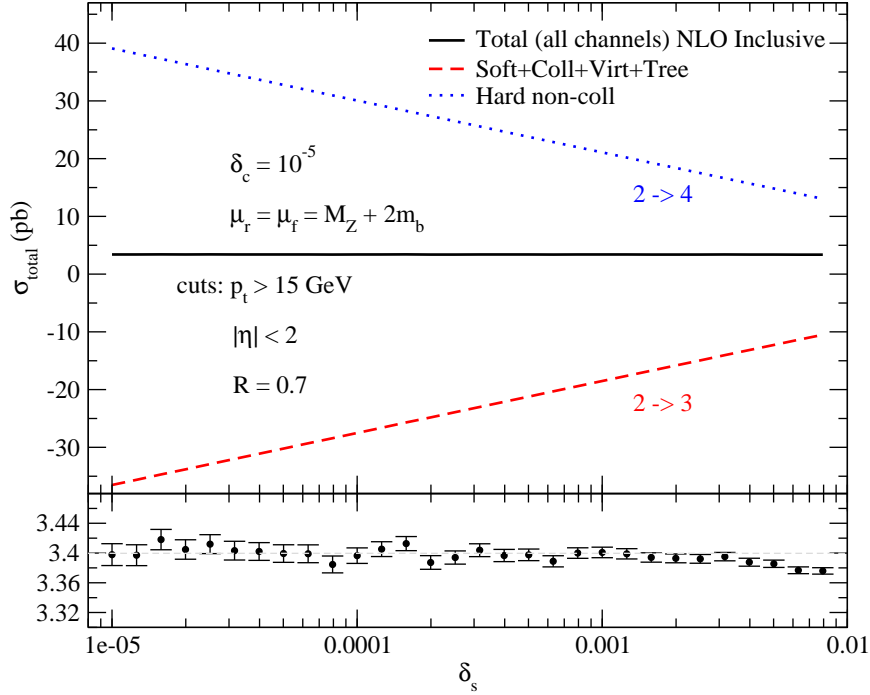}
\caption[Dependence of the total $\Zbb$ NLO QCD cross section on the
  $\delta_s$ PSS parameter.]{Dependence of $\sigma^{\rm \sss NLO}(p\bar p\to \Zbb)$ on the
  soft cutoff $\delta_s$ of the two-cutoff PSS method for
  $\mu\!=\!2m_b+M_{\sss Z}$, and $\delta_c\!=\!10^{-5}$. The upper plot
  shows the cancellation of the $\delta_s$-dependence between
  $\sigma^{soft}+\sigma^{hard/coll}$ and
  $\sigma^{hard/non-coll}$. The lower plot shows, on an enlarged
  scale, the dependence of the full $\sigma^{\rm \sss NLO}=\sigma^{\rm \sss
    NLO}_{gg}+\sigma^{\rm \sss NLO}_{q\bar{q}}+\sigma^{\rm \sss NLO}_{qg}$ on
  $\delta_s$ with the corresponding statistical errors of the MC integration.}
\label{fig:ds_dependenceZbb}
\end{center}
\end{figure}
\begin{figure}[ht]
\begin{center}
\includegraphics*[scale=0.55]{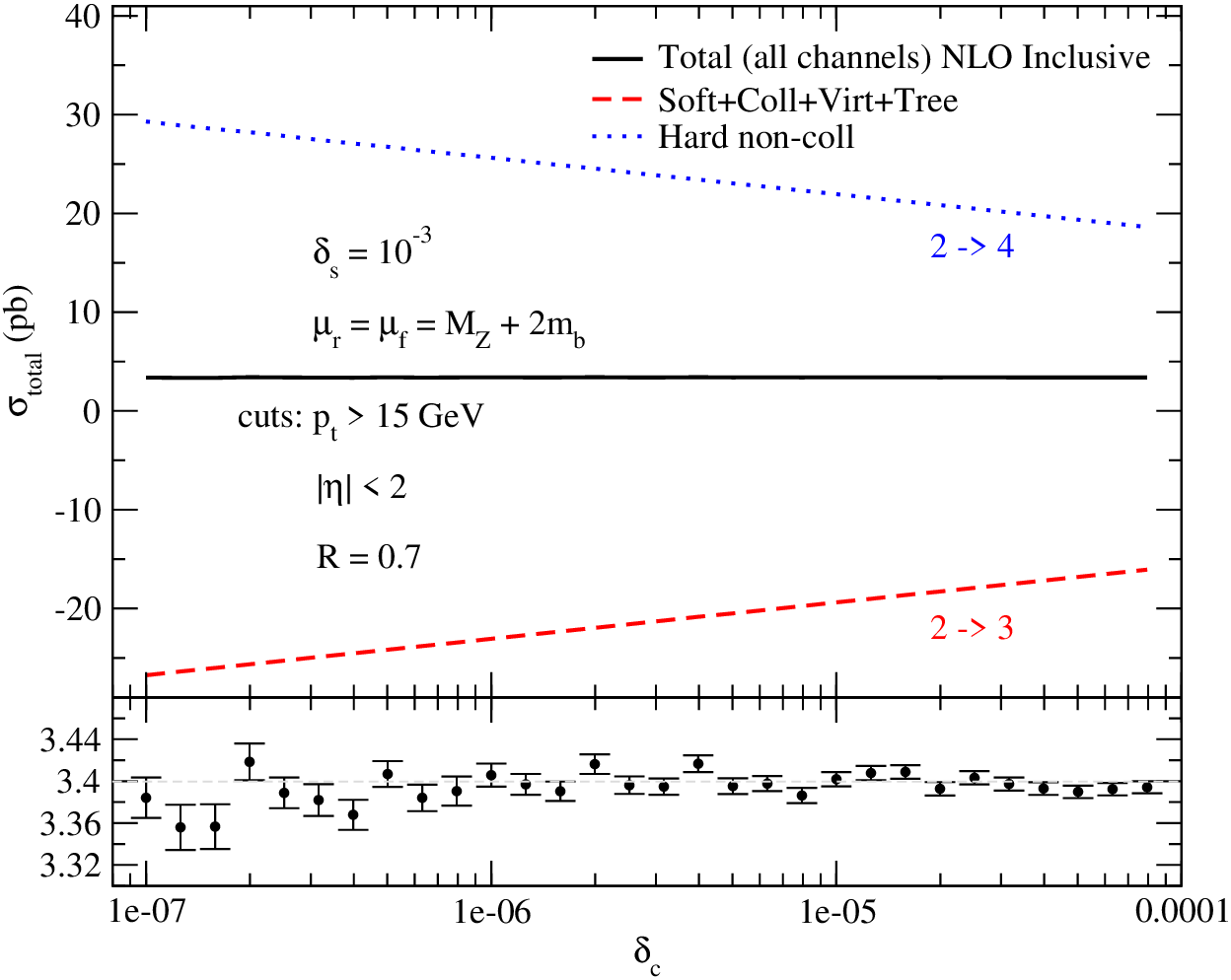}
\caption[Dependence of the total $\Zbb$ NLO QCD cross section on the
  $\delta_c$ PSS parameter.]{Dependence of $\sigma^{\rm \sss NLO}(p\bar p\to \Zbb)$ on the
  collinear cutoff $\delta_c$ of the two-cutoff PSS method, 
  for $\mu\!=\!2m_b+M_{\sss Z}$, and $\delta_s\!=\!10^{-3}$. The upper plot
  shows the cancellation of the $\delta_s$-dependence between
  $\sigma^{soft}+\sigma^{hard/coll}$, and
  $\sigma^{hard/non-coll}$. The lower plot shows, on an enlarged
  scale, the dependence of the full $\sigma^{\rm \sss NLO}=\sigma^{\rm \sss
    NLO}_{gg}+\sigma^{\rm \sss NLO}_{q\bar{q}}+\sigma^{\rm \sss NLO}_{qg}$ on
  $\delta_c$ with the corresponding statistical errors of the MC integration.}
\label{fig:dc_dependenceZbb}
\end{center}
\end{figure}
In Figs.~\ref{fig:ds_dependenceZbb} and \ref{fig:dc_dependenceZbb},
using the setup described in Section~\ref{sec:results}, we demonstrate
the independence of $\sigma^{\rm \sss NLO}(p\bar p\to \Zbb)$ on
$\delta_s$ and $\delta_c$ separately, by varying only one of the two
cutoffs over an extended range, while the other is kept fixed. In
Fig.~\ref{fig:ds_dependenceZbb}, $\delta_s$ is varied between
$10^{-5}$ and $10^{-2}$ with $\delta_c\!=\!10^{-5}$, while in
Fig.~\ref{fig:dc_dependenceZbb}, $\delta_c$ is varied between
$10^{-7}$ and $10^{-4}$ with $\delta_s\!=\!10^{-3}$. In both plots, we
show in the upper window the overall cutoff dependence cancellation
between the hadronic cross sections $\sum_{ij}
(\sigma^{soft}_{ij}+\sigma^{hard/coll}_{ij})$ and $\sum_{ij}
\sigma^{hard/non-coll}_{ij}$ in $\sum_{ij} \sigma^{\rm real}_{ij}$,
including all channels, $gg$, $q\bar q$ and $qg$. Note that we also
take into account contributions from the LO and the virtual cross
sections which are cutoff independent.  In the lower window of the
same plots we show the full $\sigma^{\rm \sss NLO}$, including all
channels, on a scale that magnifies the details of the
cutoff-dependence cancellation. The statistical errors from the MC
phase space integration are also shown.  Both
Figs.~\ref{fig:ds_dependenceZbb} and \ref{fig:dc_dependenceZbb} show a
clear plateau over a wide range of $\delta_s$ and $\delta_c$ and the
NLO cross section is proven to be cutoff independent. The numerical
results presented in Section~\ref{sec:results} have been obtained by
using the two-cutoff PSS method with $\delta_s\!=\!10^{-3}$ and
$\delta_c\!=\!10^{-5}$.

\section{Numerical results}
\label{sec:results}
The results for $Zb\bar{b}$ observables presented in this paper are
obtained for the Tevatron $p\bar{p}$ collider at $s_H=1.96$~TeV. If
not stated otherwise, we assume a non-zero bottom-quark mass, fixed at
$m_b=4.62$~GeV. The mass of the top quark, entering in the virtual
corrections, is set to $m_t=170.9$~GeV. The $Z$-boson mass is taken to
be $M_Z=91.1876$~GeV~\cite{PDBook} and the $W$-boson mass is
calculated from $M_W=M_Z \, \cos\theta_w$ with $\sin^2\theta_w=0.223$.
We work in the electroweak $G_\mu$ input scheme and replace the fine
structure constant $\alpha(0)=e^2/(4 \pi)$ by
$\alpha(G_\mu)=\frac{\sqrt{2}}{\pi} G_{\mu} M_W^2 \sin^2\theta_w$ with
the Fermi constant $G_\mu=1.16639 \cdot 10^{-5} \, {\rm GeV}^{-2}$.
The LO results use the one-loop evolution of $\alpha_s$ and the CTEQ6L1
set of PDFs~\cite{Lai:1999wy}, with $\alpha_s^{{\rm LO}}(M_Z)=0.130$, while
the NLO results use the two-loop evolution of $\alpha_s$ and the CTEQ6M
set of PDFs, with $\alpha_s^{{\rm NLO}}(M_Z)=0.118$. In the calculation of
the parton luminosity we assume five light flavors in the initial
state.  Including the $b$-quark PDF has a negligible effect ($<0.1\%$)
on the $Z b\bar{b}$ cross section and is included to consistently
compare with MCFM.  We implement the $k_T$ jet
algorithm~\cite{Catani:1992zp,Catani:1993hr,Ellis:1993tq,Kilgore:1996sq}
with a pseudo-cone size $R=0.7$ and we recombine the parton momenta
within a jet using the so called covariant
$E$-scheme~\cite{Catani:1993hr}. We checked that our implementation of
the $k_T$ jet algorithm coincides with the one in MCFM.  We require
all events to have a $b\bar{b}$ jet pair in the final state, with a
transverse momentum larger than $15$~GeV ($p_T^{b,\bar{b}}>15$~GeV)
and a pseudorapidity that satisfies $|\eta^{b,\bar{b}}|<2$. We impose
the same $p_T$ and $|\eta|$ cuts also on the extra jet that may arise
due to hard non-collinear real emission of a parton, i.e. in the
processes $Zb\bar{b}+g$ or $Zb\bar{b}+q(\bar{q})$. This hard
non-collinear extra parton is treated either \emph{inclusively} or
\emph{exclusively}. In the \emph{inclusive} case we include both two-
and three-jet events, while in the \emph{exclusive} case we require
exactly two jets in the event. Two-jet events consist of a
bottom-quark jet pair that may also include a final-state light parton
(gluon or quark) due to the applied recombination procedure. Results
in the massless bottom-quark approximation have been obtained using
the MCFM code~\cite{MCFM:2004}.

In Table~\ref{tb:ZbbXsecs} we present results for the total LO and NLO
QCD $p \bar p \to Zb \bar b$ cross sections, obtained with the scale
$\mu_r=\mu_f=M_Z+2m_b$, for both our fully massive calculation and in
the massless approximation.
\begin{table}[htp]
\begin{center}
  \caption{LO and NLO total $\Zbb$ cross sections at the Tevatron for massive and
    massless bottom quarks, using $\mu_r=\mu_f=M_Z+2m_b$. The numbers
    in square brackets are the ratios of the NLO and
    LO cross sections, the so called K-factors. Statistical errors of the
MC integration amount to about 0.1\%.}
\begin{tabular}{l|cc}
\hline
Cross Section & $m_b\ne 0$ (pb) [ratio] & $m_b=0$ (pb) [ratio] \\
\hline
$\sigma^{\rm LO}$ & $2.21 [-] $ & $2.37 [-] $ \\
$\sigma^{\rm NLO}$ \emph{inclusive} & $3.40 [1.54] $& $3.64 [1.54]$ \\
$\sigma^{\rm NLO}$ \emph{exclusive} & $2.80 [1.27] $& $3.01 [1.27]$ \\
\hline
\end{tabular}
\label{tb:ZbbXsecs}
\end{center}
\end{table}
As can be seen, the NLO QCD corrections increase considerably the
total cross section, with NLO vs. LO ratios ($K$-factors) that, in
both the massive and massless bottom-quark case, amount to $K=1.54$
and $K=1.27$ for the \emph{inclusive} and \emph{exclusive} case,
respectively.  In the following we will study the impact of the NLO
QCD corrections on $Zb\bar b$ observables in more detail. Specifically
we will show examples of kinematic distributions where a global
rescaling (or $K$-factor) does not properly describe the effect of
these corrections.
\begin{figure}[h]
\begin{center}
\includegraphics*[scale=0.4]{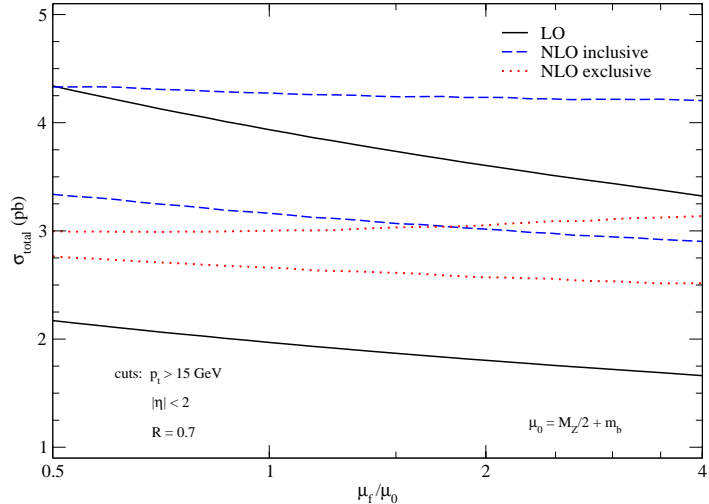} 
\caption{Dependence of the LO (black solid band), NLO
\emph{inclusive} (blue dashed band), and NLO
\emph{exclusive} (red dotted band) $\Zbb$ total cross sections on the
renormalization/factorization scales, including full bottom-quark
mass effects. The bands are obtained by independently varying both $\mu_r$ and
$\mu_f$ between $\mu_0/2$ and $4\mu_0$ (with $\mu_0=m_b+M_Z/2$).}
\label{fig:mu_dependence_band_Zbb}
\end{center}
\end{figure}
In Figs.~\ref{fig:mu_dependence_band_Zbb} and
\ref{fig:mu_dependence_Zbb} we illustrate the renormalization and
factorization scale dependence of the LO and NLO QCD total cross
sections, both in the \emph{inclusive} and \emph{exclusive} case.
Figure~\ref{fig:mu_dependence_band_Zbb} shows the overall scale
dependence of both LO, NLO \emph{inclusive} and NLO \emph{exclusive}
total cross sections, when both $\mu_r$ and $\mu_f$ are varied
independently between $\mu_0/2$ and $4\mu_0$ (with $\mu_0=m_b+M_Z/2$),
including full bottom-quark mass effects. We notice that the NLO QCD
cross sections have a reduced scale dependence over the range of
scales shown, and the \emph{exclusive} NLO QCD cross section is more
stable than the \emph{inclusive} one.  This effect is mainly driven by the
tree level subprocess $q(\bar q)g\to\Zbb+q(\bar q)$ contributing to
the real corrections. This is illustrated by the right hand side (RHS)
plots of Figs.~\ref{fig:mu_dependence_inc_Zbb} and
\ref{fig:mu_dependence_exc_Zbb}, where we show separately the
$\mu$-dependence of the total cross sections to the $q\bar{q}$,
$qg+\bar{q}g$ and $gg$ initiated processes, for $\mu_r=\mu_f$, both
for the \emph{inclusive} and for the \emph{exclusive} case.  It is
clear that the low scale behavior of the \emph{inclusive} cross
section is considerably affected by the $qg+\bar{q}g$ contribution,
which show a monotonic dependence on $\mu$ (i.e. with no plateau)
characteristic of tree level processes.  In the left hand side (LHS)
plots of Figs.~\ref{fig:mu_dependence_inc_Zbb} and
\ref{fig:mu_dependence_exc_Zbb} we also compare the scale dependence
of our results to the scale dependence of the corresponding results
obtained with $m_b=0$ (using MCFM), both at LO and at NLO. Using a
non-zero value of $m_b$ is expected to have a small impact on the
scale dependence of the results\footnote{Note that we always use
  $m_b=4.62$~GeV in the determination of the scales in terms of
  $\mu_0=m_b+M_Z/2$ even in the results obtained with $m_b=0$.}, since
the only modification to the renormalization scale dependence
originates from the bottom-quark mass and field renormalization, as
discussed in Section IIB of Ref.~\cite{Dawson:2003kb}, where we
compare the minimal and on-shell subtraction schemes.  Indeed, as can
be seen in
Figs.~\ref{fig:mu_dependence_inc_Zbb},~\ref{fig:mu_dependence_exc_Zbb}
the scale dependence of the LO and NLO curves is very similar for both
the case of a massive and massless bottom quark.
\begin{figure}[hp]
 \begin{center}
  \subfigure[\emph{Inclusive} case]{\label{fig:mu_dependence_inc_Zbb}\includegraphics*[scale=0.7]{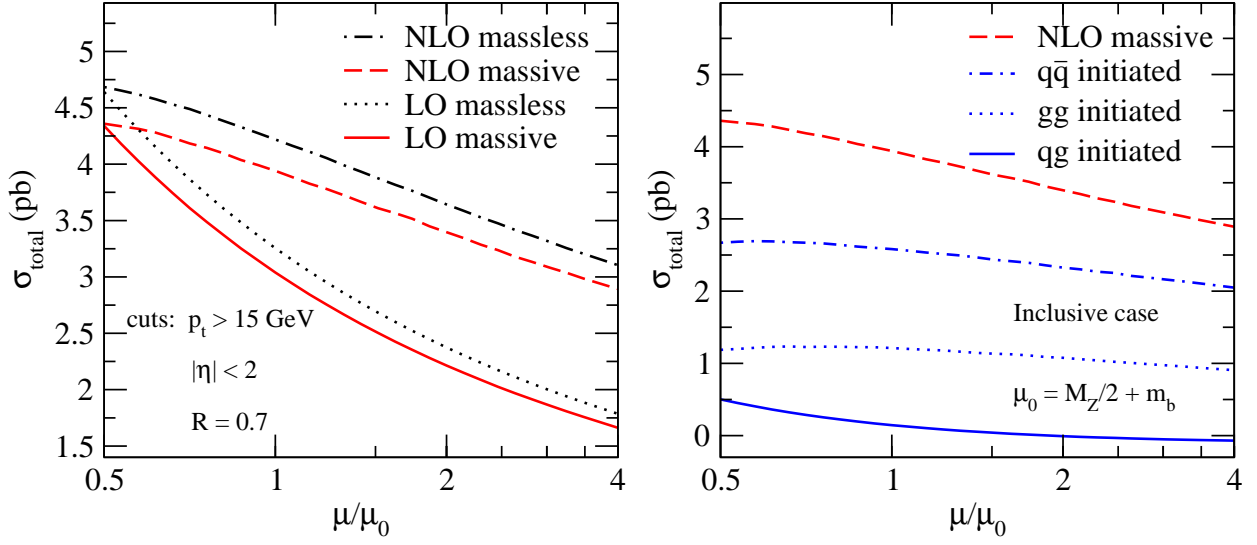}}\\
  \vspace{0.8cm}
  \subfigure[\emph{Exclusive} case]{\label{fig:mu_dependence_exc_Zbb}\includegraphics*[scale=0.7]{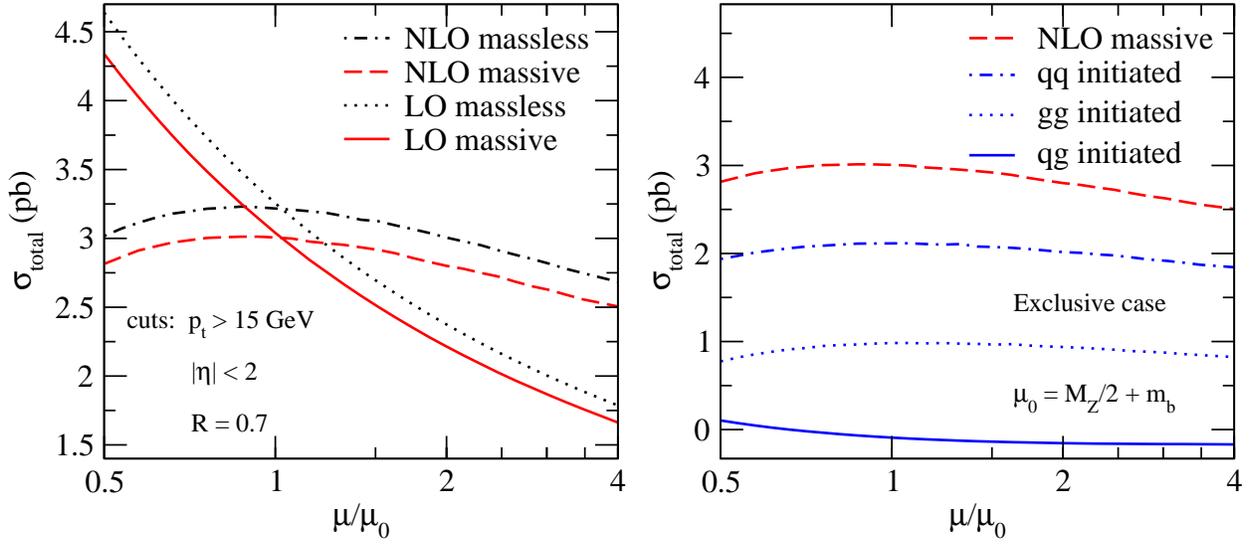}}
 \end{center}
 \caption{Dependence of the LO and NLO \emph{inclusive} and
   \emph{exclusive} $\Zbb$ total cross section on the
   renormalization/factorization scale, when $\mu_r=\mu_f=\mu$. The
   LHS plots compare both LO and NLO total cross sections for the case
   in which the bottom quark is treated as massless (MCFM) or massive
   (our calculation).  The RHS plots show separately, for the massive
   case only, the scale dependence of the $q\bar{q},gg$ and
   $qg+\bar{q}g$ contributions, as well as their sum.}
\label{fig:mu_dependence_Zbb}
\end{figure}
\begin{figure}[ht]
\begin{center}
\includegraphics*[scale=0.5]{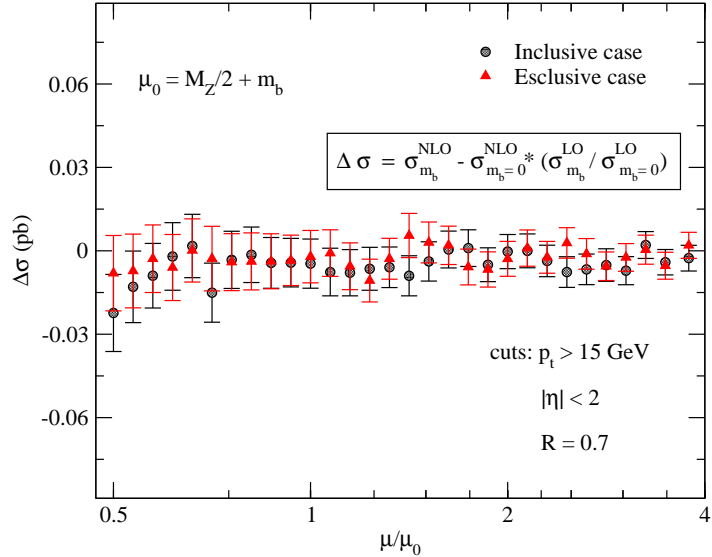} 
\caption{Dependence on the renormalization/factorization scale of the
  rescaled difference between our NLO calculation (with $m_b\ne 0$) of
  the $\Zbb$ total cross section and MCFM (with $m_b=0$) for the
  \emph{inclusive} and \emph{exclusive} cases (with
  $\mu_r\!=\!\mu_f$). The error bars indicate the statistical
  uncertainty of the MC integration.}
\label{fig:sigma_ratio_NLO_Zbb}
\end{center}
\end{figure}
While the LO cross section still has a 45\% uncertainty due to scale
dependence, this uncertainty is reduced at NLO to about 20\% for the
\emph{inclusive} and to about 11\% for the \emph{exclusive} cross
sections. The uncertainties have been estimated as the
positive/negative deviation with respect to the mid-point of the bands
plotted in Fig.~\ref{fig:mu_dependence_band_Zbb}, where each band
range is defined by the minimum and maximum value in the band.  We
notice incidentally that the difference in the total cross section due
to finite bottom-quark mass effects is less significant than the
theoretical uncertainty due to the residual scale dependence in the
\emph{inclusive} case, but is comparable in size in the
\emph{exclusive} case.  Indeed, the finite bottom-quark mass effects
amount to a reduction of the total cross sections by about 7\%
compared to the massless case at both LO and NLO QCD.

\begin{figure}[htp]
 \begin{center}
  \subfigure[\emph{Inclusive} case]{\label{fig:mbb_dist_LO_vs_NLO_inc_Zbb}
	\hspace{-0.5cm}\includegraphics*[scale=0.7]{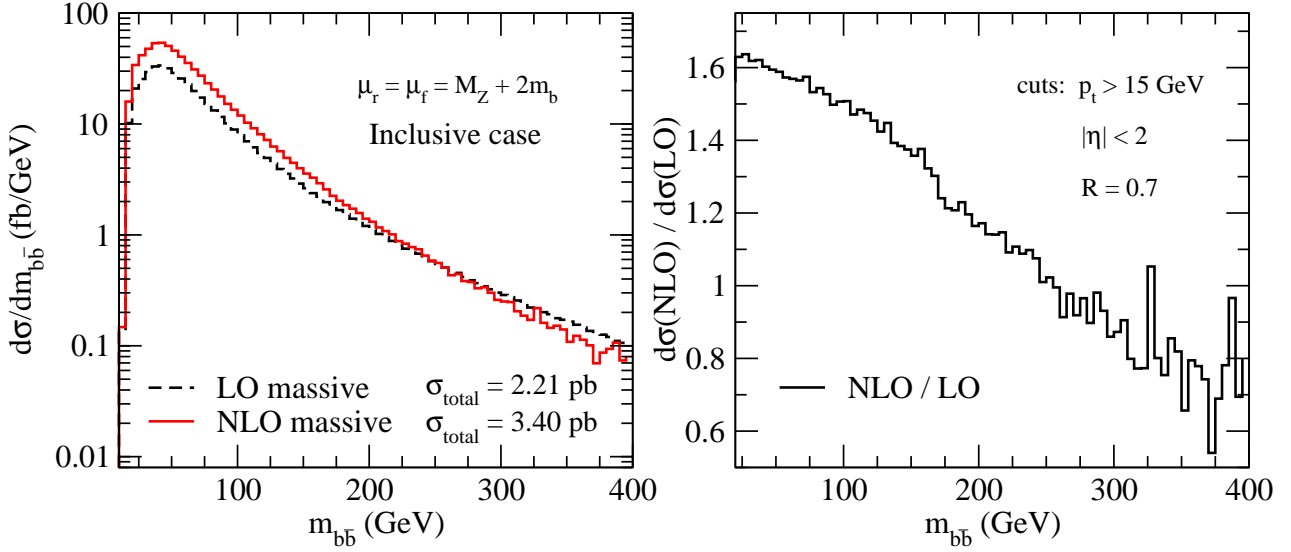}}\\
  \vspace{0.8cm}
  \subfigure[\emph{Exclusive} case]{\label{fig:mbb_dist_LO_vs_NLO_exc_Zbb}
	\hspace{-0.5cm}\includegraphics*[scale=0.7]{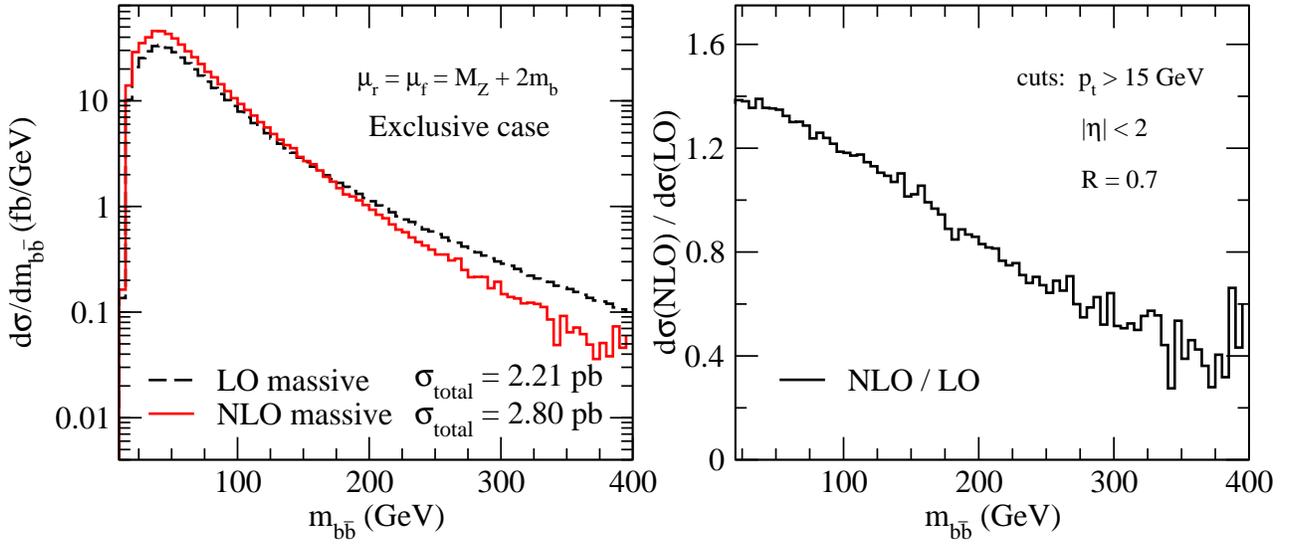}}
 \end{center}
   \caption{The distribution $d\sigma/dm_{b\bar{b}}$
      at LO and NLO QCD. The RHS plots show the ratio of
      the LO and NLO distributions.}
    \label{fig:mbb_dist_LO_vs_NLO_Zbb}
\end{figure}

In Fig.~\ref{fig:sigma_ratio_NLO_Zbb}, we show the rescaled difference between
the NLO total cross sections obtained from our calculation (with $m_b\ne
0$) and with MCFM (with $m_b=0$) defined as follows:
\begin{equation}
\Delta\sigma=\sigma^{\rm NLO}(m_b\ne 0)-\sigma^{\rm NLO}(m_b=0) 
\; \frac{\sigma^{\rm LO}(m_b\ne 0)}{\sigma^{\rm LO}(m_b=0)} \; .
\label{eq:mXsecrescaling}
\end{equation}
As can be seen, within the statistical errors of the MC integration,
the finite bottom-quark mass effects on the total cross sections at
NLO are well described by the corresponding effects at LO. 

\begin{figure}[hp]
 \begin{center}
  \subfigure[\emph{Inclusive} case]{\label{fig:mbb_dist_NLO_inc_Zbb}
	\hspace{-0.5cm}\includegraphics*[scale=0.7]{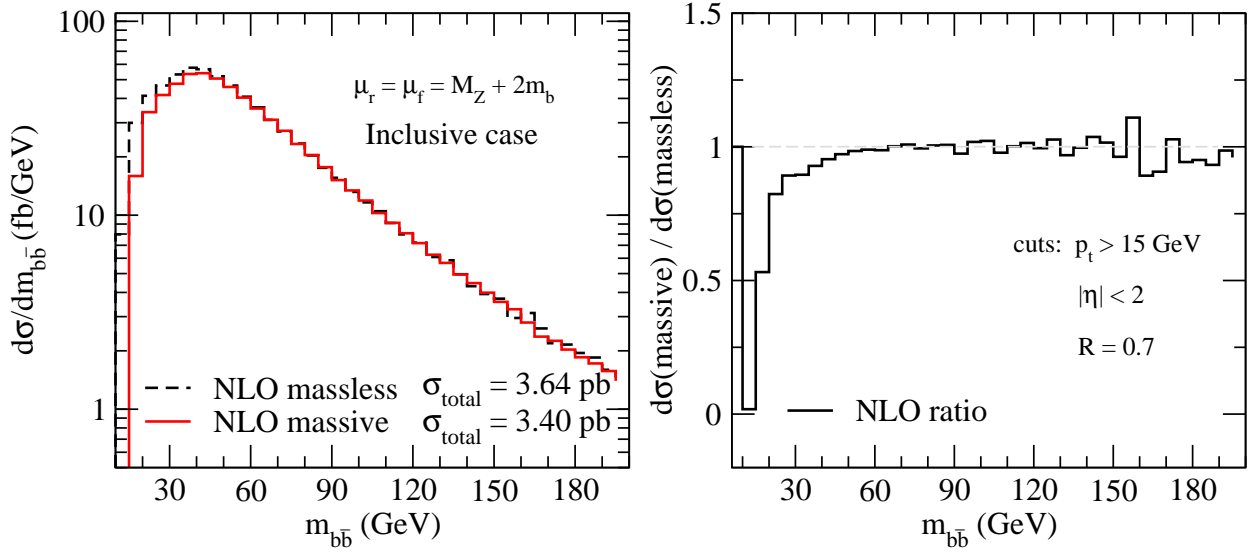}}\\
  \vspace{0.8cm}
  \subfigure[\emph{Exclusive} case]{\label{fig:mbb_dist_NLO_exc_Zbb}
	\hspace{-0.5cm}\includegraphics*[scale=0.7]{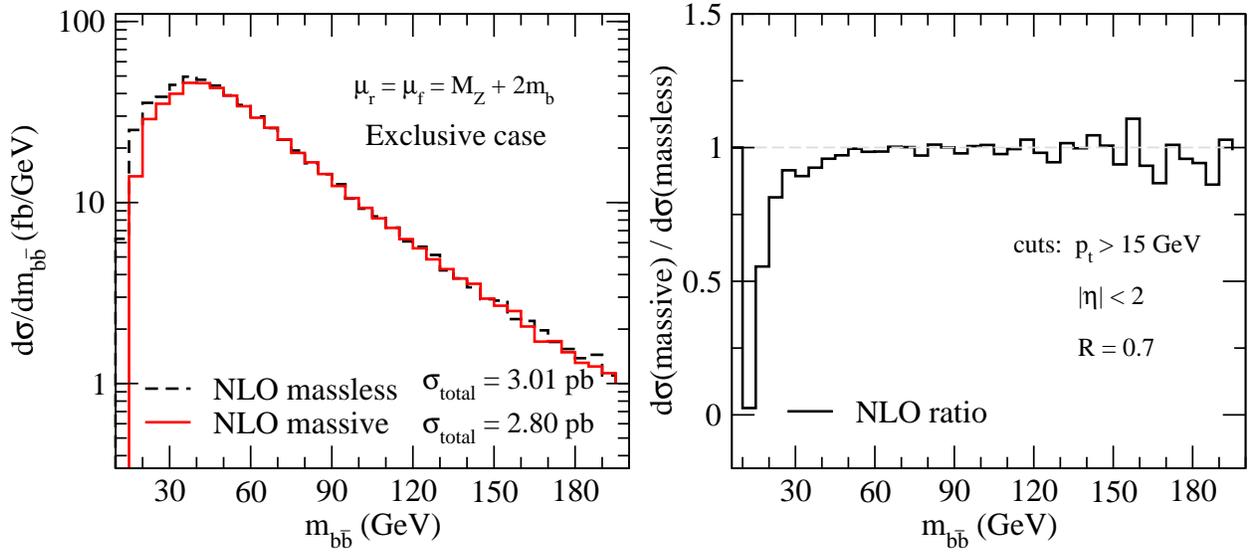}}
 \end{center}
\caption{The \emph{inclusive} and \emph{exclusive} NLO QCD distributions $d\sigma/dm_{b\bar{b}}$
derived from our calculation (with $m_b\ne 0$) and from MCFM (with
$m_b=0$).  The RHS plots show the ratio of the two
distributions, $d\sigma(m_b\neq 0)/d\sigma(m_b=0)$.}
\label{fig:mbb_dist_NLO_Zbb}
\end{figure}

\begin{figure}[hp]
\begin{center}
\includegraphics*[scale=0.65]{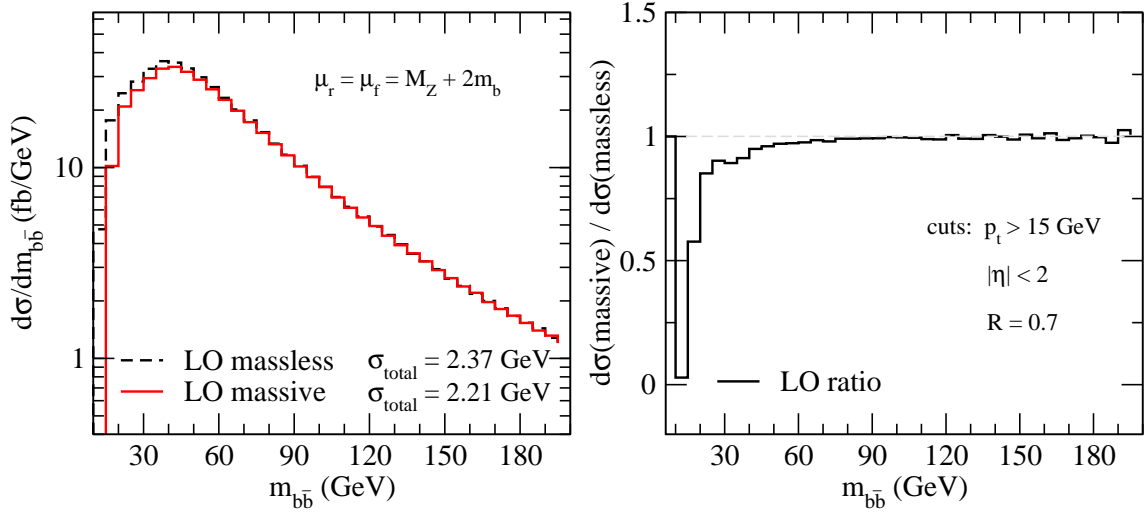} 
\caption{The LO distribution $d\sigma/dm_{b\bar{b}}$
derived from our calculation (with $m_b\ne 0$) and from MCFM (with
$m_b=0$). The RHS plot shows the ratio of the two
distributions, $d\sigma(m_b\neq 0)/d\sigma(m_b=0)$.}
\label{fig:mbb_dist_LO_Zbb}
\end{center}
\end{figure}

Finally, in Figs.~\ref{fig:mbb_dist_LO_vs_NLO_Zbb} to
\ref{fig:mbb_dist_LO_Zbb} we study the distribution
$d\sigma/dm_{b\bar{b}}$, where $m_{b\bar{b}}$ is the invariant mass of
the $b\bar{b}$ jet pair. The impact of NLO QCD corrections on this
distribution is illustrated in
Figs.~\ref{fig:mbb_dist_LO_vs_NLO_inc_Zbb} and
\ref{fig:mbb_dist_LO_vs_NLO_exc_Zbb} for the \emph{inclusive} and
\emph{exclusive} case, respectively. We see that the NLO QCD
corrections affect the differential cross section quite substantially.
In each figure the RHS plot gives the ratio of the NLO and LO
distributions.  We stress the fact that the NLO $m_{b\bar b}$ distributions
cannot be obtained from the LO ones by just rescaling, which is clear from the RHS plots of
Fig.~\ref{fig:mbb_dist_LO_vs_NLO_Zbb}.

\begin{figure}[ht]
\begin{center}
\includegraphics*[scale=0.48]{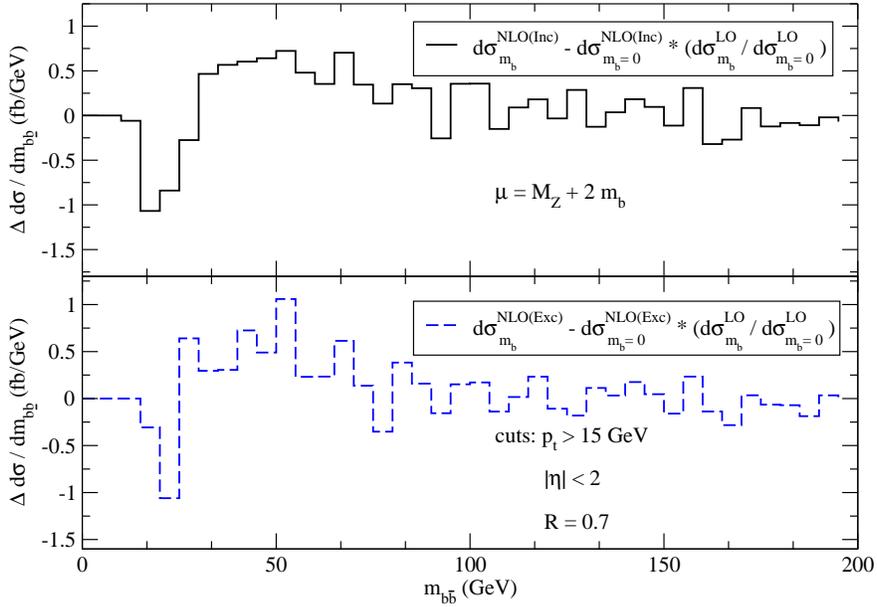} 
\caption{The $m_{b\bar b}$ distribution of the rescaled difference 
between our NLO calculation (with $m_b\ne 0$) and MCFM (with
$m_b=0$) for the \emph{inclusive} (upper plot) and \emph{exclusive} case (lower plot).}
\label{fig:mbb_ratio_NLO_Zbb}
\end{center}
\end{figure}

Figs.~\ref{fig:mbb_dist_NLO_inc_Zbb} and
\ref{fig:mbb_dist_NLO_exc_Zbb} compare the NLO $d\sigma/dm_{b\bar{b}}$
distributions obtained from the massive and massless bottom-quark
calculations. The results with $m_b=0$ have been obtained using MCFM.
As expected, most of the difference between the massless and massive
bottom-quark cross sections is coming from the region of low
$m_{b\bar{b}}$ invariant mass, both for the \emph{inclusive} and
\emph{exclusive} case, where the cross sections for $m_b\ne 0$ are
consistently below the ones with $m_b=0$.  This is emphasized in the
RHS plots, where we show the ratio of the two distributions,
$d\sigma(m_b\neq 0)/d\sigma(m_b=0)$.  For completeness, we also show
in Fig.~\ref{fig:mbb_dist_LO_Zbb} the comparison between massive
($m_b\neq 0$) and massless ($m_b=0$) calculations at LO in QCD. The LO
$m_{b\bar{b}}$ distribution for massive bottom-quarks has been
obtained both from our calculation and from MCFM, which implements the
$m_b\neq 0$ option at tree level, and both results agree perfectly.
In general, mass effects are similar at LO and NLO.  To illustrate
this in more detail we show in Fig.~\ref{fig:mbb_ratio_NLO_Zbb} the
rescaled difference between the $m_{b\bar b}$ distributions obtained
with our NLO calculation (with $m_b\ne 0$) and with MCFM (with
$m_b=0$) defined as follows:
\begin{equation}
\Delta \frac{d\sigma}{d m_{b \bar b}}=\frac{d\sigma^{\rm NLO}}{d m_{b \bar b}}(m_b\ne 0)
-\frac{d\sigma^{\rm NLO}}{d m_{b \bar b}} (m_b=0)\; \frac{d\sigma^{\rm LO}(m_b\ne 0)}{d\sigma^{\rm LO}(m_b=0)} \; .
\label{eq:mDistrescaling}
\end{equation}
We notice that, in the $\Zbb$ case, finite bottom-quark mass effects
are relevant up to values of the $m_{b\bar b}$ invariant mass around
50 GeV. Although not included in the present analysis,
our calculation is still valid when both $b$ quarks are in the forward
direction. In this region, collinear singularities can arise, which
are regularized by the finite $b$-quark mass.  The resummation of the
corresponding large logarithms is then appropriate and is left to
future improvements.

\section{Conclusions}
\label{sec:conclusions}
We have calculated the NLO QCD corrections to hadronic $Zb\bar{b}$
production including full bottom-quark mass effects.  We have
presented numerical results for the total cross section and the
invariant mass distribution of the bottom-quark pair at the Tevatron
for both massless (with MCFM) and massive bottom quarks.  We apply the
$k_T$ jet algorithm, require two $b$-tagged jets and impose
kinematical cuts that are inspired by the D$\emptyset$ and CDF
searches for the SM Higgs boson in $ZH$ production.  The NLO QCD
$Zb\bar{b}$ cross section shows a considerably reduced renormalization
and factorization scale dependence, i.~e.~about 20\% for the
\emph{inclusive} and about 11\% for the \emph{exclusive} cross
sections as opposed to 45\% scale uncertainty of the LO cross section.
The bottom-quark mass effects amount to about 7\% of the total NLO QCD
cross section and can impact the shape of the $m_{b\bar b}$
distributions, in particular in regions of low $m_{b\bar b}$. This is
relevant to SM Higgs searches in $ZH$ associated production with $H\to
b\bar{b}$ and to searches for MSSM Higgs bosons in $H b \bar{b}$
production with $H\to\mu^+\mu^-,\tau^+\tau^-$. 
We also plan to apply
the formalism developed in this paper to the calculation of both
$Zt\bar{t}$~\cite{Lazopoulos:2007bv,Lazopoulos:2008de} and $\gamma
t\bar{t}$ production at NLO in QCD. Both processes are of interest to
the study of electroweak properties of the top
quark~\cite{Baur:2004uw,Baur:2005wi}, while $Zt\bar{t}$ also
constitutes a relevant background to new physics searches.

\section*{Acknowledgements}

F.~F.~C. thanks Zvi Bern and Harald Ita for helpful discussions.  The
work of F.~F.~C. and L.~R.~is supported in part by the U.S. Department
of Energy under grants DE-FG03-91ER40662 and DE-FG02-97IR41022
respectively.  The work of D.~W.~is supported in part by the National
Science Foundation under grants NSF-PHY-0456681 and NSF-PHY-0547564.

\bibliography{zbb_nlo}

\end{document}